\documentclass[twocolumn,aps,pra,amssymb,showpacs,longbibliography,10pt]{revtex4-1}

\usepackage[normalem]{ulem}

\usepackage{amsfonts,amssymb,amsmath,graphicx,color,hyperref}

\newcommand{\vq}{\mathbf{r}}
\newcommand*\dint{\mathop{}\!\mathrm{d}}  
\newcommand{\iu}{{i\mkern1mu}} 	
\newcommand{\me}{\mathrm{e}} 	
\newcommand{\ie}{{\it i.e.~}} 	
\newcommand{\eg}{{\it e.g.~}} 	

\newcommand{\vk}{\mathbf{k}}

\newcommand{\vv}{\mathbf{v}}

\newcommand{\vsigma}{\boldsymbol{\sigma}}

\newcommand{\br}{\mathbf{r}}

\newcommand{\real}{\mathrm{Re}}
\newcommand{\imag}{\mathrm{Im}}

\newcommand{\imp}{\text{imp}}

\newcommand{\ZB}{\text{ZB}}
\newcommand{\RWA}{\text{RWA}}
\newcommand{\HDF}{\text{HDF}}
\newcommand{\vop}{\mathbf{v}} 
\newcommand{\echo}{\text{echo}}

\newcommand{\crit}{\text{crit}}

\newcommand{\tmax}{\text{max}}
\newcommand{\tmin}{\text{min}}

\newcommand{\initial}{\text{initial}}
\newcommand{\revival}{\text{revival}}

\newcommand{\st}{\text{st}}
\newcommand\aaa[1]{{\scriptsize #1}}
\newcommand{\RM}[1]{\MakeUppercase{\romannumeral #1{}}} 

\newcommand\pan[1]{{\bf({#1})}}

\definecolor{greenPR}{rgb}{0.00, 0.6, 0.00}
\newcommand{\commentout}[1]{}

\def\be{\begin{equation}}
\def\ee{\end{equation}}
\def\ber{\begin{eqnarray}}
\def\eer{\end{eqnarray}}

\graphicspath{{./pictures/}}

\begin{document}

\title{Steering Zitterbewegung in driven Dirac systems -- from persistent modes to echoes}
\author{Phillipp Reck$^{1}$, Cosimo Gorini$^1$, Klaus Richter$^{1}$}
\affiliation{$^1$Institut f\"ur Theoretische Physik, Universit\"at Regensburg, 93040 Regensburg, Germany}
\date{October 22, 2019}

\begin{abstract}
Although zitterbewegung -- the jittery motion of relativistic particles -- is known since 1930 and was predicted 
in solid state systems long ago, it has been directly measured so far only in so-called quantum simulators, 
\ie quantum systems under strong control such as trapped ions and Bose-Einstein condensates.
A reason for the lack of further experimental evidence is the transient nature of wave packet zitterbewegung.
Here we study how the jittery motion can be manipulated in Dirac systems via time-dependent potentials, 
with the goal of slowing down/preventing its decay, or of generating its revival.
For the harmonic driving of a mass term, we find persistent zitterbewegung modes in pristine,
\ie scattering free, systems.  
Furthermore, an effective time-reversal protocol
-- the ``Dirac quantum time mirror'' -- is shown to retrieve zitterbewegung through echoes.
\end{abstract}

\maketitle

%


\section{Introduction}
\label{sec_intro}

Zitterbewegung (ZB), \ie the trembling motion of relativistic particles described by the Dirac equation, 
was found by Schr\"odinger already in 1930 \cite{schroedinger1930,zawadzki2011}. 
The jittery movement is due to the fact that the velocity operator does not commute with the Hamiltonian,
and therefore it is not a constant of motion.
Indeed, the superposition of particle- and antiparticle-like solutions of the Dirac equation leads to 
harmonic oscillations with, in case of electrons and positrons, frequency $f=2mc^2/h \sim 10^{20}\,$Hz and amplitude 
given by the Compton wave length $\lambda_C\sim10^{-13}\,$m, whose direct measurement is 
still beyond experimental capabilities \cite{huang1952}.

On the other hand, the requirements for ZB are not unique to the relativistic Dirac equation, 
but can in principle be fulfilled in any two- (or multi-) band system.
Examples thereof are solid state systems with spin-orbit coupling, 
as proposed by Schliemann et al. in \RM{3}-\RM{5} semiconductor quantum wells \cite{schliemann2005,schliemann2006}, 
where the energy spectrum is formally similar to the Dirac Hamiltonian. 
In a solid state system, the ZB is directly induced  by the periodic underlying lattice \cite{zawadzki2010}. 
ZB in systems with low-energy effective Dirac-like dispersion was later proposed 
for carbon nanotubes \cite{zawadzki2006}, graphene \cite{katsnelson2006b,rusin2008} 
and topological insulators \cite{shi2013}.  Recently, ZB was further predicted for magnons \cite{wang2017} and exciton-polaritons \cite{sedov2018}.
ZB signatures can also be found in the presence of magnetic fields, \eg in graphene \cite{schliemann2008b-graph,rusin2008} 
and \RM{3}-\RM{5} semiconductor quantum wells with spin-orbit coupling \cite{schliemann2008a-2deg}.

The first experimental observations of ZB were achieved with a single $^{40}$Ca$^+$-ion in a 
linear Paul trap \cite{gerritsma2010} and for Bose-Einstein condensates \cite{qu2013,leblanc2013} 
with an induced spin-orbit coupling, using atom-light interactions \cite{wang2010}.

Recently, indirect experimental realizations of ZB in solid state systems were also reported \cite{stepanov2016, iwasaki2017}.
Also motivated by these, we study time-dependent protocols aimed at prolongating the ZB duration
or at generating revivals by effectively time-reversing its decay. 
To the best of our knowledge there are currently very few studies of ZB in time-dependent driving fields.
In one case graphene in an external monochromatic electromagnetic field was considered,
and multimode ZB, \ie ZB with additional emerging frequencies, was obtained but found to decay over
time \cite{rusin2013}.  
Another work suggests that time-dependent Rashba spin-orbit coupling
in a two-dimensional electron gas might indefinitely sustain ZB \cite{cong2014}.

Our goal is two-fold: (i) to identify non-decaying ZB modes in driven Dirac systems, \eg graphene;
(ii) to consider the possibility of generating ZB ``echoes'' exploiting the time-mirror protocols 
put forward in \cite{reck2017,reck2018b}.  We start in Sec.~\ref{sec:gen-ZB} with a succinct introduction to ZB
in a static Dirac system, highlighting the mechanisms leading to ZB decay and laying out our general strategy to counteract it
via different drivings.  
In Sec.~\ref{sec:timedep-ZB} we show that a monochromatic time-modulation 
of the mass term yields multimode ZB with more frequencies as compared to driving from a monochromatic electromagnetic field, where most importantly additional modes turn out to be long-lived.
Section~\ref{sec:echo} deals with the generation of ZB echoes/revivals, which instead require 
a short mass gap pulse.  Section~\ref{sec:concludeZB} concludes.

\section{Zitterbewegung in Dirac systems: frequency, amplitudes and decay}
\label{sec:gen-ZB}

Consider the static Dirac Hamiltonian
\begin{equation}
 H_0 = \hbar v_F \vk\cdot \vsigma + M_0 \sigma_z. 
\label{eq:Hamil-tindepZB}
\end{equation}
Its eigenenergies are
\begin{equation}
\varepsilon_\pm(\vk) = \pm\sqrt{M_0^2+\hbar^2 v_F^2 |\vk|^2 } = \pm M_0 \sqrt{1+\kappa^2} 
\label{eq:epsZB}
\end{equation}
where 
\begin{equation}
\kappa = \hbar v_F |\vk|/M_0.
\label{eq:kappa}
\end{equation}
The eigenstates are
\begin{equation}
 | \varphi_{\vk,\pm}\rangle =  \frac{1}{\sqrt{2}\sqrt{1+\kappa^2\pm\sqrt{1+\kappa^2}}} \begin{pmatrix}
                                          1\pm\sqrt{1+\kappa^2} \\ \kappa \,\me^{\iu \gamma_\vk}
                                         \end{pmatrix} \, | \vk\rangle,
\label{eq:HchiZB}
\end{equation}
with $\gamma_\vk$ being the azimuthal angle of $\vk$ measured from the $k_x$-axis.

Considering an initial plane wave with wave vector $\vk$ living in the two bands,
\begin{equation}
 |\psi_{0,\vk}\rangle = c_\vk^+ |\varphi_{\vk,+}\rangle + c_\vk^-|\varphi_{\vk,-}\rangle,
\end{equation}
its time evolution is trivially given by
\begin{equation}
 |\psi_\vk(t)\rangle = c_\vk^+ \me^{-\iu\omega_{\vk,+} t} |\varphi_{\vk,+}\rangle + c_\vk^- \me^{-\iu\omega_{\vk,-} t} |\varphi_{\vk,-}\rangle,
\label{eq:psi_k(t)}
\end{equation}
with eigenfrequencies $\hbar \omega_{\vk,\pm}=\varepsilon_{\pm}(\vk)$.
The ZB is generated by the interference term in the time-dependent expectation value of the velocity operator
\begin{equation}
  \langle \vop^\ZB \rangle_\vk (t) = 2\operatorname{Re}\lbrace c_\vk^+ (c_\vk^-)^\ast \, \me^{-\iu \Omega^\st_\vk t } \, \langle \varphi_{\vk,-}\mid\vop\mid\varphi_{\vk,+}\rangle \rbrace   \label{eq:generalZB}.
\end{equation}
Here, $\vop = \nabla_\vk H_0/\hbar = v_F \vsigma$ is the velocity operator and the ZB frequency is given by
\begin{equation}
 \hbar\Omega_\vk^\st = \varepsilon_+(\vk) - \varepsilon_-(\vk) = 2 M_0 \sqrt{1+\kappa^2},
\label{eq:ZB-freq-gappedgraph}
\end{equation}
where ``st'' stands for ``static''.
Evaluating the matrix element of the velocity operator for the gapped Dirac system yields 
both parallel and perpendicular ZB, with amplitudes
\begin{align}
 A^{\st,\parallel}_\vk =& \, v_F \frac{2|c_\vk^+||c_\vk^-|}{\sqrt{1+\kappa^2}} \leq \frac{v_F }{\sqrt{1+\kappa^2}}, \\
 A^{\st,\perp}_\vk =&\, v_F 2|c_\vk^+||c_\vk^-| \,\leq v_F.
\end{align}
In the perfect (scattering-free) system described by $H_0$, ZB of a single $\vk$-mode
oscillates without decaying, with an amplitude and frequency given by the initial band structure occupation. 
On the contrary, ZB of a wave packet has a transient character \cite{lock1979}, \ie it vanishes over time. 
For an initial wave packet of the general form
\begin{equation}
 |\psi_0\rangle = \int \dint^2k\, \psi_0(\vk) \, |\psi_{0,\vk}\rangle,
\end{equation}
one has
\begin{equation}
  \langle \vop^\ZB \rangle (t) = 
\int \dint^2k\, |\psi_0(\vk)|^2\, \langle \vop^\ZB\rangle_\vk (t),  \label{eq:generalZB-WP1}
\end{equation}
\ie the wave packet ZB is the average of the plane wave ZB weighted by the $\vk$-space distribution of the initial state.
As different $\vk$-modes have different frequencies, such a collective ZB dephases over time and vanishes.
Technically, this is due to the phase $\me^{-\iu \Omega^\st_\vk t }$ in Eq.~\eqref{eq:generalZB},
whose oscillations as a function of $\vk$ become faster for increasing time $t$ -- and thus average progressively to zero.
Rusin and Zawadzki give an alternative but equivalent explanation for the ZB decay \cite{rusin2007}. 
They start by considering the movement of the two sub-wave packets in the different $\pm$ bands,
each made up of modes with velocities $\vv_\pm = \nabla_\vk \varepsilon_\pm(\vk)/\hbar$. 
Since $\vv_+$ is antiparallel to $\vv_-$, the sub-packets move away from each other
and progressively decrease their mutual overlap, which translates to a decrease of the interference and thus of ZB.

This paper is devoted to circumventing or reverting the decay of the ZB via a time-modulation of 
the mass term $M_0\rightarrow M_0 + M(t)$.  More precisely, we consider the general time-dependent Hamiltonian  
\begin{equation}
H = \hbar v_F \vk\cdot \vsigma + M_0 \sigma_z + M(t) \sigma_z = H_0 + H_1(t)
\label{eq:Hamil-tdepZB}
\end{equation}
and study two scenarios.
The first one is based on harmonic (monochromatic) driving, to be dealt with in Sec.~\ref{sec:timedep-ZB}.  
Here we follow the strategy of Ref.~[\onlinecite{rusin2013}],
determining analytically the emerging frequencies of the driven ZB in our system
via the rotating wave approximation (RWA) and the high-driving frequency (HDF) limit. 
Our analytics are then compared to numerical simulations based on the 
``Time-dependent Quantum Transport" (TQT) software package \cite{krueckl2013},
which also allows us to study multimode ZB in regimes not accessible analytically.
By taking a Fourier transform of the numerically obtained time-dependent velocity 
we can identify the oscillation frequencies, and aim at finding out long-lived or 
possibly non-decaying modes -- \ie we are after oscillations which survive on a long time scale.
To single-out the large time oscillations we Fourier transform the simulation signal for $t>t^*$, 
where $t^*$ is the time when the amplitude of the initial transient oscillations has
decayed below 5\% of its $T=0$ value, see Sec.~\ref{subsec:longt-ZB}.

The second scenario, discussed in Sec.~\ref{sec:echo}, is radically different:
rather than looking for long-lived modes in response to a persistent, monochromatic driving,
we consider the effects of a sudden (non-adiabatic) on-and-off modulation of the gap.
The idea is to use the quantum time mirror protocol of Refs.~\cite{reck2017,reck2018b}
to effectively time-reverse the sub-wave packet dynamics.  Once the latter are brought back together
we expect a reconstruction of the interference pattern yielding ZB.

\section{Driven zitterbewegung in Dirac systems: Emergence of persistent multimodes}
\label{sec:timedep-ZB}
We start from Eq.~\eqref{eq:Hamil-tdepZB} with a harmonically oscillating mass term of the form
\begin{equation}
 M(t) = \tilde M \cos(\omega_D t)
\label{eq:Mharmonic}
\end{equation}
and study the resulting ZB, \ie we time evolve a given initial state according to 
\begin{equation}
 \iu \hbar \frac{\partial}{\partial t} \psi(t) = H(t) \psi(t) 
\label{eq:tSE}
\end{equation}
and calculate the expectation value of the velocity operator.

\subsection{Driven Zitterbewegung: Analytics}
In the following analytical section, we adapt a procedure from Ref.~[\onlinecite{rusin2013}], details of the derivation can be found in Ref.~[\onlinecite{reckphd}].  
\subsubsection{Rotating wave approximation (RWA)}
\label{subsec:RWA}


The rotating wave approximation (RWA) is well-known and often used in quantum optics 
to simplify the treatment of the interaction between atoms, \ie few-level systems, and a laser field.
Although here we consider two \emph{bands}, the system is effectively 
a two-\emph{level} system for any arbitrary $\vk$ as long as $\vk$ is conserved, \ie for homogeneous pulses.
The conditions for the applicability of the RWA are: (i) the amplitude of the time-dependent part, $\tilde M$,
has to be small compared to other internal energy scales of the system; (ii) its frequency is in resonance with one of the level spacings: $\omega_D\approx\Omega^\st_\vk$.
In that case, all high-frequency terms in the Hamiltonian average out at physical time scales 
and only the resonant terms survive \cite{parisibook}.
Calculations are done for a single $\vk$-mode, since the collective wave packet ZB is given 
by the weighted superposition from Eq.~\eqref{eq:generalZB-WP1}.

The goal is to solve the time-dependent Dirac equation \eqref{eq:Hamil-tdepZB}, 
\ie to find for all $\vk$ the time-dependent occupations of its two nonperturbed bands, given certain initial conditions.
One first looks for an SU(2) (pseudo)spin rotation which diagonalizes the time-dependent part of the Hamiltonian. 
Then, according to the RWA, only slow terms are kept, \ie outright static ones or those whose time-dependence 
is given by $\me^{\pm\iu(\omega_D-\Omega^\st_\vk)t}$. 
Faster terms, in our case with a time-dependence $\me^{\pm\iu(\omega_D+\Omega^\st_\vk)t}$ or $\me^{\pm\iu\omega_Dt}$, 
average out rapidly and are dropped.
The equations then decouple, yielding a homogeneous second order differential equation
of the harmonic oscillator type.  Its time-dependent wave function is then used to compute the expectation 
value of the velocity operator $\vop$.
The ZB perpendicular to the propagation direction $\vk$ yields
\begin{align}
 \langle v_\perp \rangle_\vk =& v_F \frac{\hbar\sqrt{1+\kappa^2}}{\tilde M \kappa} \Big\lbrace   |A_+|^2  (\Delta +\omega_R) \sin\left((\omega_D+\omega_R) t\right) \nonumber \\
& +  |A_-|^2  (\Delta -\omega_R)\sin\left((\omega_D-\omega_R) t\right) \nonumber \\
&+  2 \omega_R  \,  \imag \left( A_-^\ast A_+ \me^{-\iu \omega_D t}\right) 
\Big\rbrace,
\label{eq:ZB-vperpRWA}
\end{align}
where we define two additional characteristic frequencies
\begin{align}
 \Delta &= \Omega^\st_\vk - \omega_D, \\
 \label{omegaR}
 \omega_R &= \sqrt{\Delta^2 + \frac{\tilde M^2}{\hbar^2} \,\frac{\kappa^2}{1+\kappa^2} }.
\end{align}
The quantities $A_\pm$ are given by the initial conditions (occupation of the two bands) as shown in Appendix \ref{app:A+A-}.

The perpendicular ZB oscillates with three distinct frequencies: $\omega_D$, $\omega_D\pm \omega_R$.
This is in contrast to the standard electromagnetic driving scenario, where only the two
frequencies $\omega_D\pm\omega_R$ are obtained \cite{rusin2013}.  Crucially, the additional $\omega_D$-mode
turns out to be non-decaying and thus determining the ZB long-time behavior as discussed in Sec.~\ref{subsec:longt-ZB}.
Although not shown here, the static limit can be derived from Eq.~\eqref{eq:ZB-vperpRWA}, 
\eg by taking the limit $\tilde M\to0$.  

Similar features arise for the parallel-to-$\vk$ ZB component 
\begin{align}
 \langle v_\parallel \rangle_\vk = \frac{\hbar v_F}{\tilde M \kappa} &\Big\lbrace   |A_+|^2  (\Delta +\omega_R) \cos\left((\omega_D+\omega_R) t\right)  \nonumber \\
&+  |A_-|^2  (\Delta -\omega_R) \cos\left( (\omega_D-\omega_R) t\right) \nonumber \\
&+ 2 \Delta \,  \real\left( A_+^\ast A_- \me^{\iu \omega_D t}\right) \Big\rbrace \nonumber \\
&+  4v_F\cos(\omega_R t)\real\left\lbrace A_+^\ast A_- \right\rbrace + \text{const}.
\label{eq:ZB-vparallelRWA}
\end{align}
There are now four different frequencies, $\omega_D, \omega_D\pm\omega_R$ and $\omega_R$,
as opposed to the single $\omega_R$-mode of electromagnetic driving 
\footnote{This is very likely due to the fact that Ref.~[\onlinecite{rusin2013}] considers pristine graphene, 
where no parallel ZB is expected in the static case.}.
The $\omega_D$-mode is once again responsible for the long-time behavior.


\subsubsection{High driving frequency (HDF)}
\label{subsec:HDF}
%

We now investigate the ZB for high driving frequencies (HDF) $\hbar \omega_D \gg \tilde M$, thus
extending the analytically accessible regions.  The derivation is similar to RWA with a different initial
$SU(2)$ transformation \cite{rusin2013}.
The HDF approximation, $\me^{\pm 2 \iu \frac{\tilde M}{\hbar\omega_D} \sin(\omega_D t)}\approx 1$, 
allows for solving the remaining differential equations analytically. 
One obtains (for perpendicular ZB)
\begin{align}
 \langle v_\perp\rangle_\vk = &-\frac{1}{k} \left( \text{const} +  2 \Omega_\vk^\st \imag \left\lbrace B_+ B_-^\ast \me^{\iu\Omega^\st_\vk t} \right\rbrace \right) \nonumber \\
				  & - \frac{4\,\tilde M}{k \hbar\omega_D} \frac{M_0}\hbar \, \sin(\omega_D t) \, \real \left\lbrace B_+ B_-^\ast \me^{\iu\Omega^\st_\vk t} \right\rbrace,
\end{align}
and (for parallel ZB)
\begin{align}
 \langle v_\parallel\rangle_\vk = &-\frac{1}{k} \left( \text{const} +  4\frac{M_0}\hbar \real \left\lbrace B_+ B_-^\ast \me^{\iu\Omega^\st_\vk t} \right\rbrace \right) \nonumber \\
				  &+ \frac{2\,\tilde M}{k \hbar\omega_D} \Omega_\vk^\st \, \sin(\omega_D t) \, \imag \left\lbrace B_+ B_-^\ast \me^{\iu\Omega^\st_\vk t} \right\rbrace,
\label{eq:HDF-parallel}
\end{align}
where the quantities $B_\pm$ are again given by the initial conditions, see Appendix \ref{app:A+A-}.
Both $ \langle v_\perp\rangle_\vk$ and $\langle v_\parallel\rangle_\vk$ have an $\Omega^\st_\vk$-mode as in the static case, 
as well as weaker $\mathcal{O}(\tilde M / \hbar\omega_D)$ oscillations at 
$\omega_D \pm  \Omega^\st_\vk$. 
Their suppression, going along with the survival of the $\Omega^\st_\vk$-mode, has a simple physical reason:
Electrons cannot respond to a driving much faster than the frequencies of their intrinsic dynamics and therefore oscillate at $\Omega^\st_\vk$ as if no extra field was present.

\subsection{Driven Zitterbewegung: Numerics}
\label{subsec:num-ZB-tdep}

\begin{figure}
 \centering
    \def\svgwidth{\columnwidth}
    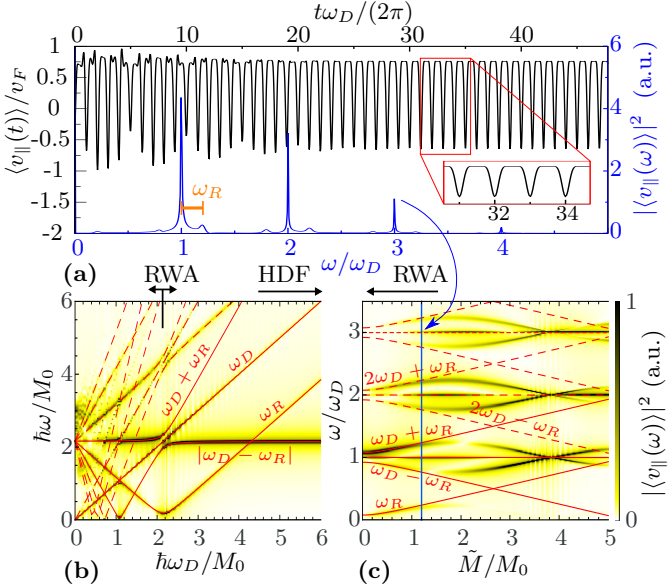
 \centering 
\caption{
Frequencies of driven zitterbewegung (ZB).
 (a)  The parallel ZB  $\langle v_\parallel(t)\rangle$ is shown both, as a function of time (black) and its Fourier transform (blue), \ie as function of $\omega$, for $\omega_D = 2M_0/\hbar$ and $\tilde M = 1.2 M_0$. 
The inset shows a close-up for larger times. Peaks appear in the Fourier transform at multiple integers of $\omega_D$ with satellite peaks at a distance of $\omega_R$ (orange line at the first peak) away from the major peaks.
(b) and (c): Fourier transform of $\langle v_\parallel\rangle$ as a function of driving frequency $\omega_D$ (for fixed $\tilde M = 0.5M_0$) and amplitude $\tilde M$ (for fixed $\omega_D= 2M_0/\hbar$) of the time-dependent mass term, respectively.
The blue vertical line in (c) marks the function shown in panel (a) (indicated by the blue arrow). 
The expected frequencies from RWA, $\omega_D$, $\omega_D\pm\omega_R$ and $\omega_R$ are shown (red, solid), as well as higher-order terms in $\omega_D$ (red, dashed). 
For smaller $\tilde M$, up to $\tilde M \approx 1.3M_0$ the RWA results, Eq.~\eqref{eq:ZB-vparallelRWA}, are well recovered while for larger $\tilde M$, the RWA is not justified anymore.
In all plots, we choose $\vk_0$ such that $\varepsilon_+(\vk_0) = 0.4 M_0$, and thus the ZB frequency corresponding to the static $H_0$ is $\Omega^\st_{\vk_0} \approx 2.15 M_0/\hbar$.
 } \label{fig:ZB-FT}
\end{figure}

 
We test our RWA and HDF analytical results against numerical simulations
based on the TQT wave packet propagation package \cite{krueckl2013}, which uses a Lanczos method \cite{lanczos1950} to evaluate the action of the
time-evolution operator on a wave packet to time evolve an initial state numerically.

In the following, the initial state is a Gaussian wave packet
\begin{equation}
 \psi_0(\vk)= \frac{1}{\sqrt{\pi\Delta k^2}} \exp\left( -\frac{(\vk-\vk_0)^2}{2\Delta k^2} \right),
\end{equation}
that equally occupies both bands, and which is time-evolved in the presence of the time-dependent mass potential.
If not otherwise specified, we take $M_0$ as the energy unit choose the center wave vector $\vk_0$ such that $\hbar v_F k_0 = 0.4 M_0$ and a $\vk$-space width $\Delta k  =  |\vk_0|/10$, where $k_0=|\vk_0|$.
The velocity expectation value is obtained by numerically computing the
time-derivative of the wave packet position expectation value
\begin{equation}
\langle \vop(t_i) \rangle = \frac{\langle\br(t_{i+1})\rangle - \langle\br(t_{i})\rangle}{\delta t},
\end{equation}
at each time $t_i$ with $\delta t$ the simulation time step.
The static ZB frequency is $\Omega^\st_{\vk_0} \approx 2.15 M_0/\hbar$.

The upper trace in Fig.~\ref{fig:ZB-FT}(a) shows the simulation data for $\langle v_\parallel \rangle$ as a function of time (black),
for large-amplitude driving, $\tilde M = 1.2 M_0$, at frequency $\omega_D = 2M_0/\hbar$.
After an initial irregular transient ($t\omega_D/2\pi \lesssim 20$) a stable periodic signal settles. 
Note however that the long-time response is not monochromatic, as manifest from the zoom-in inset.
We discuss this in detail below in Sec.~\ref{subsec:longt-ZB}.

The numerically computed fast Fourier transform of $\langle v_\parallel(t) \rangle$ 
is shown in Fig.~\ref{fig:ZB-FT}(a) (blue).
Clear peaks are visible at integer multiples $\omega=n\,\omega_D,\;n\in\mathbb{N}$, 
accompanied by smaller satellite peaks at $n\omega_D\pm\omega_R$.
To visualize the dependence on the different parameters, 
the fast Fourier transform $|\langle v_\parallel(\omega) \rangle|^2$ is shown in density plots,  
each vertical slice corresponding to one simulation at a given value of the driving frequency $\omega_D$ for a fixed amplitude $\tilde M = 0.5 M_0$ [panel (b)] and at a given value of the driving amplitude $\tilde M$ for a fixed frequency $\omega_D = 2M_0/\hbar$ [panel (c)].
%
The vertical blue line in panel (c) corresponds to the simulation shown in panel (a), 
as indicated by the blue arrow.  The ZB behaves as expected from HDF, Eq.~\eqref{eq:HDF-parallel}, for large $\omega_D$ (only mode: $\Omega^\st_{\vk_0}$). 
Furthermore, in the region of validity of RWA, the frequencies $\omega_D$, $\omega_D\pm\omega_R$ and $\omega_R$ are present [solid red lines in panels (b) and (c)]
as long as $\tilde{M} \lesssim M_0$ and $\omega_D \approx \Omega^\st_{\vk_0}$.
Note that for $\hbar\omega_D\gg \tilde M$ one has $\omega_R  \xrightarrow{\omega_D\to\infty} \omega_D - \Omega^\st_{\vk_0}$,
so that one of the RWA modes becomes $|\omega_D - \omega_R| \xrightarrow{\omega_D\to\infty} \Omega^\st_{\vk_0}$ and
coincides with the HDF result.  Indeed, the line labeled $|\omega_D - \omega_R|$ in panel (b) 
merges with the RWA modes in the region $\omega_D \approx \Omega^\st_{\vk_0}$.

In addition to the expected frequencies, further modes also emerge, 
obtained by adding integer multiples of $\omega_D$ to the lower ones.
The reason for the appearance of the latter is given in Appendix \ref{app:n*om_D},
based on higher-order time-dependent perturbation theory.

\subsection{Long-time behavior of the zitterbewegung}
\label{subsec:longt-ZB}

\begin{figure}
 \centering 
    \def\svgwidth{\columnwidth}
    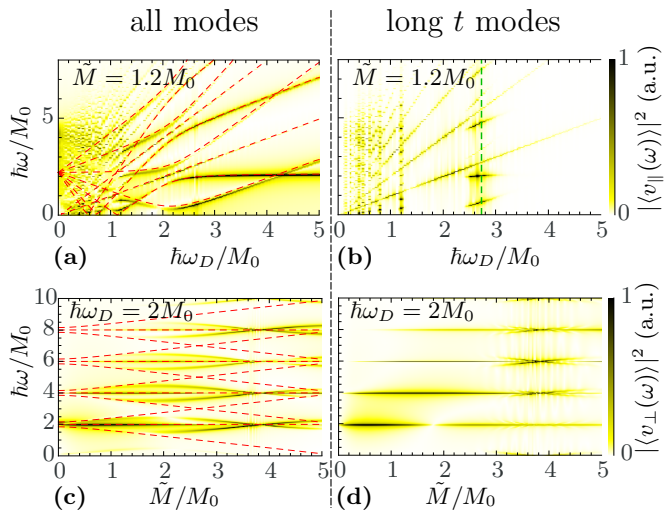
 \centering 
\caption{
 Searching for infinitely long surviving modes of the zitterbewegung. 
The left panels (a), (c) show plots of all appearing ZB modes whereas the right panels (b), (d) show the modes which survive a long time (as defined in the text).  
The modes of the perpendicular  and parallel 
ZB $\langle v_\perp (\omega) \rangle $ (upper panels) and $\langle v_\parallel (\omega) \rangle $ (lower panels) are shown as a function either of the parameter $\omega_D$ or $\tilde M$. 
Analytically expected modes from RWA (red, dashed) are indicated in the left panels. 
In general, the surviving modes for long times are the ones not depending on $\vk$, \eg multiples of $\omega_D$, or weakly $\vk$-dependent where $\partial \omega_R/\partial k = 0$ [indicated by the green dashed lines in (b)].
  \label{fig:ZB-FT-longt}}
\end{figure}

We now search for long-lived ZB modes.  If present, they should easier to be detected experimentally.
The long-time ZB frequencies are easily obtained from the simulation data:
by considering the Fourier transform of the signal starting at a time $t>t^*$ after
the initial transient has decayed.
The time $t^*$ is defined by the condition that the relative ZB amplitude in the time-\emph{in}dependent setup 
has decreased to less than 5\% in both $\langle v_\parallel \rangle$ and $\langle v_\perp \rangle$ (if both are present). 

In Fig.~\ref{fig:ZB-FT-longt}, the ZB frequency spectra for a Fourier transform starting at $t=0$ (left panels) 
and for a Fourier transform starting at $t^*$ (right panels) are compared for several parameter combinations. 
The simulation data is the same on both sides, only the time interval for the fast Fourier transform changes:
$[0, t_\text{max}]$ to identify all modes, or $[t^*, t_\text{max}]$ to single-out the long-lived ones, where $t_\text{max}$ is maximal time of the simulation.
%
Comparing the left and right panels in Fig.~\ref{fig:ZB-FT-longt}, it is clear that
some branches fade out completely, some remain nearly unchanged, and others still survive only in a small parameter regime.
This can be understood by recalling the general arguments from Sec.~\ref{sec:gen-ZB}, 
where the ZB decay was shown to be due to dephasing caused by the varying ZB frequencies of different $\vk$-modes building the propagating wave packet.
Therefore ZB modes that weakly depend on $\vk$ -- or are 
outright $\vk$-independent -- will not dephase and should thus survive.

The RWA and HDF approximations indicate that the only $\vk$-independent ZB mode has frequency $\omega_D$. 
Indeed, in all plots modes with frequencies $\omega_D$ and integer multiples thereof 
are unchanged in the long-time limit.
The strangely regular shape of the timeline for the long-lived ZB in the close-up of Fig.~\ref{fig:ZB-FT}(a) 
might be explained by the fact that mostly integer multiples of $\omega_D$ survive. 
This corresponds to a discrete Fourier transform and thus to a $\omega_D$-periodic behavior in time.
These $\vk$-independent modes are the only truly infinitely-lived ones.
However, they might be expected, since it is not too surprising that a system
driven at $\omega_D$ will respond at the same frequency (and at multiples thereof).
More interestingly, Fig.~\ref{fig:ZB-FT-longt} (b) shows that sections of certain branches survive 
even at frequencies which are not integer multiples of $\omega_D$.
Such modes are locally $\vk$-independent, \ie the ZB frequencies are stationary with respect to changes in $\vk$. 
Using Eq.~\eqref{omegaR} for the RWA frequency $\omega_R$ and setting $\partial \omega_R /\partial k \stackrel{!}{=} 0$ yields
\begin{align}
& \Delta \frac{\partial \Omega^\st_\vk}{\partial{k}} + v_F \frac{\tilde M^2}{\hbar} \frac{\kappa}{(1+\kappa^2)^2} = 0 
\nonumber \\
\Leftrightarrow& -M_0\omega_D + \Omega^\st_\vk M_0 + \frac{\tilde M^2}{2\sqrt{1+\kappa^2}^3} = 0.
\label{eq:domR/dk}
\end{align}
To analyze the data in Fig.~\ref{fig:ZB-FT-longt}(b), where the ZB is shown as a function of $\omega_D$, 
we solve Eq.~\eqref{eq:domR/dk} for $\omega_D$:
\begin{equation}
\omega^\crit_D = \Omega^\st_\vk \left(1 + \frac{\tilde M^2}{4M_0^2} \frac{1}{(1+\kappa^2)^2}\right).
\end{equation}
The result is shown as green dashed line in Fig.~\ref{fig:ZB-FT-longt}(b),
confirming that the system also responds (persistently) at frequencies not directly related to that of the driving.
Different ZB modes depend on $\vk$ via $\omega_R$, and indeed each branch intersecting the green line
has long-lived components around the intersection point.
The long-lived modes are huddled around the exact value $\omega^\crit_D\simeq2.75 M_0/\hbar$ 
(for the given simulation parameters).  
Notice that additional surviving modes, \ie (locally) $\vk$-independent, 
appear at small frequencies, \eg $\omega_D = 1.2 M_0/\hbar$.
The latter cannot be explained within RWA, which looses validity for $\omega_D < \Omega^\st_\vk = 2.15\,M_0/\hbar$.

The last kind of modes which survive for longer times can be seen in panels (c) and (d) at $\tilde{M}\approx 4M_0$, where the $n\omega_D$-modes are crossed by other modes. 
However, their frequency is close to the surviving $n\omega_D$-modes and therefore they do not particularly alter the long-time behavior, which is why we relocate the numerical discussion of their appearance to Appendix \ref{app:crossing_survive}.

\section{Zitterbewegung echoes via effective time-reversal}
\label{sec:echo}

\begin{figure*}
 \centering
    \def\svgwidth{\textwidth}
    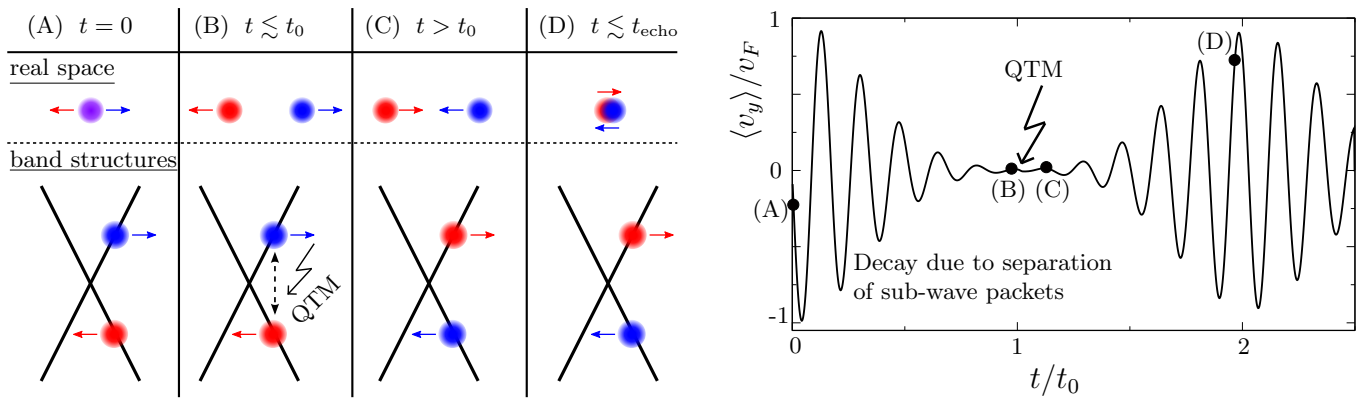
 \centering 
\caption{
Retrieving decaying zitterbewegung through a spin-echo type time-reversal mechanism. 
Left panel: An initial wave packet with components in both bands 
separates due to the different band velocities $\propto\nabla_\vk \varepsilon(\vk)$. 
The decreasing overlap between the sub-wave packets causes the ZB to decay [(A) to (B)], see the timeline of the velocity expectation value $\langle v_y\rangle$ in the right panel. 
Via a short ''quantum time mirror`` (QTM) pulse at $t_0$ the sub-wave packets switch bands, thus reversing their velocities
and returning to their initial positions [see (C)].
The recovery of the overlap in (D) leads to a ZB revival.
Black dots in the right panel belong to the time instants sketched in the left panel.
 } \label{fig:ZBecho-mechanism}
\end{figure*}

We now discuss an alternative way to retrieve late-time information of ZB.
Instead of Eq.~\eqref{eq:Mharmonic}, we consider a short mass pulse of the form
\begin{equation}
\label{eq:Mstep}
 M(t) = 
 \left\{
 \begin{array}{cl}
 \tilde M, & \quad t_0 < t < t_0 + \Delta t, \\
 0, & \quad {\rm otherwise.} 
 \end{array}
 \right.
\end{equation}
The step-like form \eqref{eq:Mstep} is chosen for definiteness and mathematical convenience only.
Indeed, as long as the mass pulse is diabatically switched on and off it can act as a Quantum Time Mirror (QTM), 
\ie it can effectively time-reverse the dynamics of a wave packet, 
irrespectively of its detailed time profile \cite{reck2018a,reck2018b}.
In general, we expect that effective time-reversal will yield an echo of the initial ZB.
This is apparent when recalling that a propagating two-band wave packet progressively splits into
two sub-wave packets, each composed of states belonging to one of the two bands,
and that ZB is due to the interference between the sub-wave packets.
As the spatial overlap between the latter decreases, so does their interference,
causing the ZB to decay \cite{rusin2007}. 
A properly tuned mass pulse, however, causes the sub-wave packets to invert their occupation of the two bands (the former electron-like state becomes a hole-like state and vice-versa), and hence also invert their direction of motion.
Thus, the sub-wave packets start to reapproach each other and as they recover their initial overlap the interference pattern yielding ZB is reconstructed,
see Fig.~\ref{fig:ZBecho-mechanism}.

\subsection{ZB echoes: Analytics}

We closely follow Refs.~\cite{reck2017,reck2018b}, and quantify the strength of the ZB echo
by considering the density correlator
\begin{equation}
 \mathcal{C}(t) = \int \dint^2r \, |\psi(\br,t)| \,|\psi(\br,0)|,
 \label{eq:corr}
\end{equation}
\ie the spatial overlap between the initial density and the one at time $t$.
This measure is appropriate for wave packets which are initially well localized in space.
The mass pulse action on an initial eigenstate is $\vk$-conserving (the pulse is homogeneous in space) 
and can in general be expressed as a change of the initial band occupation
\begin{equation}
 |\psi_{\vk,s}\rangle = B_s(\vk,\Delta t) |\varphi_{\vk,s}\rangle + A_s(\vk,\Delta t)|\varphi_{\vk,-s}\rangle.
 \label{eq:action_pulse}
\end{equation} 
For gapped Dirac systems the transition amplitude $A_s$ is independent of $s$, $A_s\rightarrow A$.
One has \cite{reck2018b}
\begin{equation}
A(\vk,\Delta t) =  \frac{i\kappa\tilde M}{\sqrt{M_0^2+M^2\kappa^2}\,\sqrt{1+\kappa^2}} \sin\left( \frac{M\Delta t}{\hbar} \sqrt{1+\kappa^2} \right),
\label{eq:transamp-hyper}
\end{equation}
with $M= M_0 + \tilde M$ and $\kappa$ from Eq.~\eqref{eq:kappa}.  Each $\vk$-mode contributes to the echo only
with its component which switches bands (reverses the velocity) during the mass pulse, 
\ie the term proportional to $A$ in Eq.~\eqref{eq:action_pulse}.
Moreover, the components proportional to $B_s$ are irrelevant and will be neglected in the following.  
Given a general initial wave packet 
\begin{equation}
\label{eq:generic_wavepacket}
 |\psi_0 \rangle = \sum\limits_\vk \psi_0(\vk)  |\psi_0^\vk \rangle  =   \sum\limits_\vk \psi_0(\vk) \left(c^+_\vk |\varphi_{\vk,+} \rangle + c^-_\vk |\varphi_{\vk,-} \rangle \right),   
\end{equation}
with $|c^+_\vk|^2 + |c^-_\vk|^2 = 1$, this amounts to considering only its effectively 
time-reversed part
\begin{align}
 |\psi_\echo (t^\prime) \rangle =&  \sum\limits_\vk \psi_0(\vk) \; |\psi_\echo^\vk (t^\prime) \rangle \nonumber \\
=& \sum\limits_\vk \psi_0(\vk) \; \sum\limits_{s=\pm 1}  c^{s}_\vk  \,\me^{- \iu \omega_{\vk,s} t_0} \,  \nonumber \\
&\times   A(\vk,\Delta t) \, \me^{-\iu \omega_{\vk,-s} t_1}\,|\varphi_{\vk,-s}\rangle.    
 \label{eq:phi_echo-t>t_0}
\end{align}
Here $\hbar\omega_{\vk,\pm s}=\varepsilon_\pm(\vk)$, see Eq.~\eqref{eq:epsZB},
while $t'=t_0 + \Delta t + t_1$ is a generic time after the pulse.
For the initial wave packet \eqref{eq:generic_wavepacket}, the ZB at time $t^\prime$ reads
\be
 \langle \vop^\ZB\rangle (t^\prime) = \int \dint^2k\, |\psi_0(\vk)|^2 \, \langle  \vop^\ZB \rangle_\vk (t^\prime).
\label{eq:v^ZBt>t0}
\ee
Before the pulse, each $\vk$-mode contribution is [see Eq.~\eqref{eq:generalZB}]
\begin{align}
\langle  v^\ZB_{i} \rangle_\vk (t<t_0) &= 2\operatorname{Re}\lbrace c_\vk^+ (c^-_\vk)^\ast \, \me^{-\iu \Omega^\st_{\vk} t } \, \langle \varphi_{\vk,-} \mid v_i \mid \varphi_{\vk,+}\rangle \rbrace    
\label{eq:v_k^ZBt<t0}
\end{align}
where $i\in \lbrace x,y\rbrace$ denotes the direction.

\begin{figure*}[t]
 \centering 
    \def\svgwidth{\textwidth}
    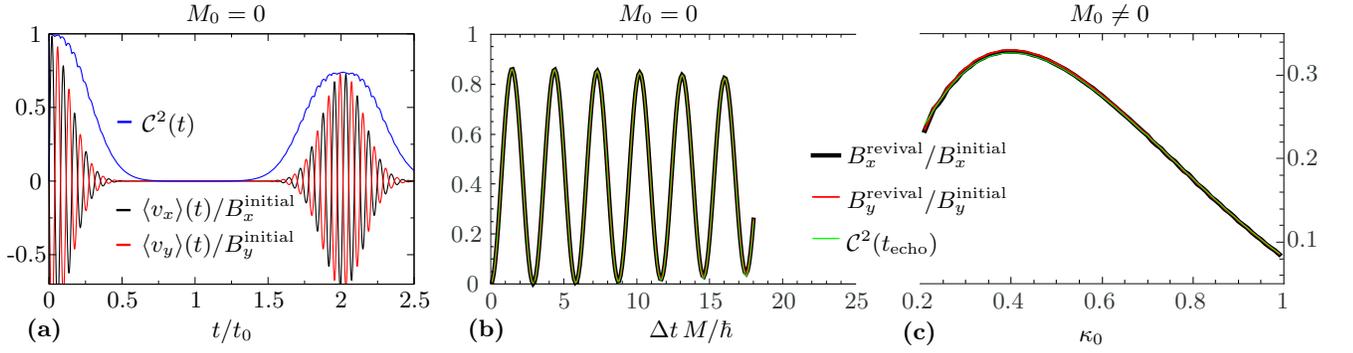
 \centering 
\caption{
Comparison between ZB echo simulations and analytical estimates for a gapless [panels (a), (b)] and a gapped 
[panel (c)] Dirac system.  
In panel (a), both the velocity expectation value and the correlation $\mathcal{C}^2(t)$, Eq.~\eqref{eq:corr}, are shown as functions of time, where $\mathcal{C}^2(t)$ and the ZB amplitude coincide at $t_\echo \simeq 2t_0$,
in agreement with Eq.~\eqref{eq:revAmpl-Corr}. 
Panel (b) compares the echo strength $\mathcal{C}(t_\echo) \approx |A(\vk_0,\Delta t)|$ (see Ref.~\cite{reck2017}), 
to the relative amplitude of the revived ZB both in $x$- and $y$-direction as a function of $\Delta t$ for
fixed $\kappa_0=\hbar v_F k_0/M=0.4$. The expected agreement can be seen.
Panel (c) shows analogous data for a gapped Dirac system, where $\Delta t=1.4 \hbar/M$
while the mean wave vector of the wave packet, \ie $\kappa_0$, is varied.
\label{fig:ZB-Revival-works}}  
\end{figure*}

Recall that the wave packet ZB decay is due to dephasing among its constituent modes: 
At larger times $t$ the exponential $\me^{-\iu \Omega^\st_{\vk} t}$ oscillates faster
as a function of $\vk$, and thus averages out in the integral \eqref{eq:v^ZBt>t0}.  
To reverse this dephasing process the phase of the oscillations must be inverted. 
%
For a time $t^\prime$ after the pulse, the velocity expectation value of each $\vk$-mode of the effectively time-inverted wave packet $|\psi_\echo\rangle$ in Eq.~\eqref{eq:phi_echo-t>t_0} yields
\begin{align}
 \langle   v^\ZB_{i} & \rangle_\vk (t^\prime) =  |A(\vk)|^2  \nonumber\\
 &\times2\real \left \lbrace  c_\vk^+ (c^-_\vk)^\ast \, \me^{-\iu \Omega^\st_{\vk} (t_0-t_1) } \, \langle \varphi_{\vk,+} \mid v_i  \mid \varphi_{\vk,-}\rangle  \right\rbrace .
\label{eq:v_k^ZBt>t0}
\end{align}
Indeed, the kinetic phase of the complex exponential $\me^{-\iu \Omega^\st_{\vk} (t_0-t_1) }$ decreases with $t_1$
(the time elapsed after the pulse) and reaches zero at $t_1 = t_0$ -- \ie the initial phase is recovered.  
The recovery happens simultaneously for all $\vk$-modes, leading to complete rephasing at the echo time
$t_\echo = 2t_0+\Delta t$.  
\footnote{At $t_\echo$ the matrix element of the velocity operator in Eqs.~\eqref{eq:v_k^ZBt<t0} and \eqref{eq:v_k^ZBt>t0} reads \unexpanded{$\langle \varphi_{\vk,+}| v_{i} | \varphi_{\vk,-} \rangle = \langle \varphi_{\vk,-} | v_{i} | \varphi_{\vk,+} \rangle^\ast$}, \ie the phase changes sign. 
For gapless Dirac systems, the matrix element \unexpanded{$\langle \varphi_{\vk,-} | \vop | \varphi_{\vk,+} \rangle$} is purely imaginary.  This gives rise to a $\pi$ phase jump, irrelevant for the ZB amplitude. However, the phase change is in general $\vk$-dependent, and its effect is discussed in Appendix \ref{app:ZBecho-phases} but neglected henceforth.}
For a wave packet centered at $\vk_0$ and narrow enough to approximate $ A(\vk) \approx A(\vk_0)$, 
the transition amplitude can be taken out of the integral (see Ref.~\cite{reck2017}) and the ratio 
of the revived ZB amplitude in direction $i$, $B_i^\revival$, to the initial one $B_i^\initial$ is
\begin{equation}
 \frac{B_i^\revival}{B_i^\initial } \approx |A(\vk_0)|^2,
\label{eq:revivedAmplitude}
\end{equation}
with $i\in \lbrace x,y\rbrace$.
On the other hand, the correlator, Eq.~\eqref{eq:corr}, is approximately 
given by the transition amplitude \cite{reck2017},
\begin{equation}
 \mathcal{C}(t_\echo) \approx |A(\vk_0,\Delta t)|,
\end{equation}
so that we expect
\begin{equation}
 \frac{B_i^\revival}{B_i^\initial } \approx \mathcal{C}^2(t_\echo)
\label{eq:revAmpl-Corr}
\end{equation}
to be checked below numerically.

As a side remark, note that the ZB echo bears a certain resemblance to the spin echo:
The latter is achieved when dephased, oscillating spins -- \ie an ensemble of two-level systems --
are made to rephase again by a $\pi$-pulse.  Here, an analogous rephasing among oscillating 
$\vk$-modes -- \ie an ensemble of delocalized states with a given dispersion -- is obtained via
the QTM protocol.

\subsection{ZB echoes: Numerics}

We numerically compute the correlator $\mathcal{C}(t)$, Eq.~\eqref{eq:corr} via TQT for gapless ($M_0 = 0$) 
and gapped ($M_0\neq 0$) Dirac systems.  
In both cases, the transition amplitude is given by Eq.~\eqref{eq:transamp-hyper}.
The initial wave packet is composed of $\vk$-modes from both branches of the Dirac spectrum,
which is necessary for ZB at $t=0$.  
As a proof of principle we use a Gaussian wave packet, narrow in reciprocal space ($\Delta k = k_0/8$), 
and compare the results with the estimates Eqs.~\eqref{eq:revivedAmplitude}-\eqref{eq:revAmpl-Corr}. 
The wave packet is peaked around $\vk_0 = (k_0,k_0)^T/\sqrt{2}$, with $\kappa_0 = \hbar v_F k_0 /M = 0.4$. 

Figure~\ref{fig:ZB-Revival-works} (a) compares the correlator $\mathcal{C}(t)$ to the relative amplitude of the revived ZB.  
The velocities are normalized with respect to the initial ZB amplitude $B_i^\initial$. 
As expected from the analytics, at the echo time $t_\echo \simeq 2 t_0$ the 
ZB amplitude coincides with the echo strength $\mathcal{C}^2(t_\echo) \simeq |A(\vk_0,\Delta t)|^2$.
Panel (b) shows the relative amplitude both in $x$- and $y$-direction, Eq.~\eqref{eq:revivedAmplitude},
and echo strengths $\mathcal{C}^2(t_\echo)$ as functions of $\Delta t$,
\ie of the pulse duration $\Delta t$. 
Numerically, the amplitude of the echo is obtained by searching for the largest difference between consecutive local maxima and minima, which are in a certain time interval around the expected echo time $t_\echo=2t_0+\Delta t$, $I=\lbrack t_\echo -\tau , t_\echo +\tau \rbrack$, to exclude the initial ZB from the automatic search in the data. 
We define the echo amplitude $B_i^\revival$ of the ZB as
\begin{equation}
 B_i^\revival = \max\limits_{t\,\in\, I} \frac{|v_i^\tmax(t)-v_i^\tmin(t+\pi/\Omega_{\vk_0}^\ZB)|}{2},
\end{equation}
where $v_i^\tmax$ is a local maximum and $v_i^\tmin$ a local minimum of the velocity in direction $i$. 
Here, we use as interval width $\tau = 0.5 t_0$, but the exact value does not matter, as long as the revival is included in the interval $I$ and the ZB that does not belong to the revival is excluded.
Since no difference between $\mathcal{C}^2$ and the ratio between initial and revived amplitude is visible in panel (b), the echo of the ZB has the expected strength, see Eq.~\eqref{eq:revAmpl-Corr}.

%
In panel (c), a gapped Dirac system ($M_0\neq 0$) is considered and the echo strengths are shown for varying mean wave vectors $\kappa_0 = \hbar v_F k_0 /M$ and fixed $\Delta t = 1.4 \hbar/M$. 
The relative amplitude of the revived ZB matches the quantum time mirror echo strength obtained by the correlation $\mathcal{C}^2(t_\echo)$, confirming again our analytical expectations.



Hence, we have shown that the ZB echo behaves as the quantum time mirror in Ref.~\cite{reck2018b} where for different band structures the effects of position-dependent potentials in the Hamiltonian, like disorder and electromagnetic fields are discussed additionally. 
To show that we recover the same results for the ZB echo also in these cases, we discuss the effect of disorder on the ZB echo in App.~\ref{app:disorderZB}.
As expected, the relative echo strength decreases exponentially in time, where the decay time is given by the elastic scattering time.

\section{Conclusions}
\label{sec:concludeZB}
We theoretically investigated different aspects of the dynamics of driven ZB via analytical and numerical methods,
within a single-particle picture.

A periodically driven mass term in (gapped) Dirac systems, \eg graphene, was shown to give rise to multimode ZB.
The latter long-time behavior reveals the existence of persistent ZB modes, emerging at frequencies
stationary with respect to wave vector changes and not necessarily at simple multiples of the driving frequency.
Such long-lived modes should allow for an experimental detection of ZB, much easier than standard, rapidly-decaying ZB modes.

Moreover, ZB revivals/echoes were shown to be generated via the QTM protocol \cite{reck2017,reck2018b},
and to behave as expected in the presence of disorder.  Though ZB experiments remain challenging \cite{stepanov2016},
it would be interesting to transfer established spin echo-based protocols -- mostly $T_2$-weighted imaging --
to ZB.

\begin{acknowledgments}
We acknowledge support from Deutsche Forschungsgemeinschaft within GRK 1570 and SFB 1277 (project A07).
\end{acknowledgments}

\appendix

\section{Relation between $A_\pm$ and $B_\pm$ and the initial occupations of a given wave packet}
\label{app:A+A-}
This appendix shows the relation between the amplitudes of a given initial state and the quantities $A_\pm$ and $B_\pm$ which are used to denote the amplitudes of the ZB in RWA and HDF, respectively.

\subsection{$A_\pm$ in rotating wave approximation}
In RWA, the ansatz used to solve the transformed Dirac equation reads 
\begin{equation}
\psi_\vk^\RWA(t) = \begin{pmatrix} A_+ \me^{-\iu \frac{\omega_D+\omega_R}{2}t}  +   A_- \me^{-\iu \frac{\omega_D-\omega_R}{2}t } \\
				 C_+ \me^{\iu \frac{\omega_D+\omega_R}{2}t}  +   C_- \me^{\iu \frac{\omega_D-\omega_R}{2}t}  \end{pmatrix},
\label{eq:psiRWA00}
\end{equation}
with 
\begin{equation}
 C_\pm = A_\pm\frac{\hbar \sqrt{1+\kappa^2}}{\tilde M \kappa }  \,\me^{\iu  \gamma_\vk}\, \left(\Delta \pm \omega_R\right),
\end{equation}
and thus for $t=0$:
\begin{equation}
\psi_\vk^\RWA(0) = \begin{pmatrix} A_+   +   A_-  \\
				 \frac{\hbar \sqrt{1+\kappa^2}}{\tilde M \kappa }  \,\me^{\iu  \gamma_\vk} \left[A_+ \left(\Delta + \omega_R\right)  +   A_-\left(\Delta - \omega_R\right) \right]\end{pmatrix}.
\end{equation}
Note that $\psi_\vk^\RWA$ solves the Dirac equation obtained after rotating the spin degree of freedom such that the static Hamiltonian is diagonal.
In other words, to compare $A_\pm$ with the amplitudes of any initial state
\begin{equation}
\phi_\vk^0 = \begin{pmatrix}\alpha_\vk \\ \beta_\vk  \end{pmatrix},
\label{eq:psi0initappendix}
\end{equation}
the transformation $S^\RWA$ is introduced through
\begin{equation}
 S^\RWA \phi_\vk^0 = \psi_\vk^\RWA(0).
\end{equation}
For the Hamiltonian in Eq.~\eqref{eq:Hamil-tdepZB}, the transformation is
\begin{equation}
 S^\RWA = \me^{-\iu\frac{\vartheta_\vk}{2} \mathbf{m}_\vk\cdot\vsigma},
\end{equation}
where $\mathbf{m}_\vk = (k_y, -k_x, 0)$ is the rotation axis and the amount of rotation $\vartheta_\vk = \left(\frac{\pi}{2} - \arctan^{-1}\frac{M_0}{\hbar v_F k}\right)$ is equal to the polar angle in the Bloch sphere.

\subsection{$B_\pm$ at high driving frequency}
In the HDF limit, the ansatz that is used to solve the Dirac equation is 
\begin{align}
 &\psi^\HDF_\vk(t) 
 = \begin{pmatrix} \left\lbrack B_+ \me^{\iu\Omega^\st_\vk t/2} + B_- \me^{-\iu\Omega^\st_\vk t/2}\right\rbrack\me^{-\iu\frac{\tilde M}{\hbar\omega_D} \sin(\omega_D t)} \\  \left\lbrack D_+ \me^{\iu\Omega^\st_\vk  t/2} + D_- \me^{-\iu\Omega^\st_\vk  t/2}\right\rbrack\me^{+\iu\frac{\tilde M}{\hbar\omega_D} \sin(\omega_D t)} \\ \end{pmatrix} \nonumber \\
\end{align}
with 
\begin{equation}
 D_\pm = - B_\pm \frac{\me^{\iu\gamma_\vk} }{2 v_F k}  \left(2\frac{M_0}\hbar\pm\Omega^\st_\vk\right)
\end{equation}
and thus for $t=0$:
\begin{equation}
\psi_\vk^\HDF(0) = \begin{pmatrix} B_+   +   B_-  \\
				-\frac{\me^{\iu\gamma_\vk} }{2 v_F k}  \left[B_+ \left(2\frac{M_0}\hbar\pm\Omega^\st_\vk\right)  +   B_-\left(2\frac{M_0}\hbar\pm\Omega^\st_\vk\right) \right]\end{pmatrix}.
\end{equation}
Since there is no transformation used in the considered Hamiltonian for the HDF case, we can directly compare $B_\pm$ with the amplitudes of any initial state of the form of Eq.~\eqref{eq:psi0initappendix}.
In contrast to RWA, we directly obtain
\begin{equation}
\phi_\vk^0 = \psi_\vk^\HDF(0).
\end{equation}

\section{Emergence of higher $\omega_D$ modes in the driven zitterbewegung}
\label{app:n*om_D}
%
%
This appendix is supposed to motivate qualitatively why for larger driving amplitudes, $\tilde M$, higher order frequencies $n \omega_D\pm\omega_R$ of the ZB appear instead of only the ZB frequencies predicted by RWA, $\omega_D\pm\omega_R$.
In the following we use the static energy eigenstates $|\varphi_{\vk,\pm}\rangle$ as basis. Any plane wave at $\vk$ can be written in that basis similar to the static case in Eq.\eqref{eq:psi_k(t)},
\begin{equation}
 |\psi_\vk(t)\rangle = c_\vk^+(t) \me^{-\iu\omega_{\vk,+} t} |\varphi_{\vk,+}\rangle + c_\vk^-(t) \me^{-\iu\omega_{\vk,-} t} |\varphi_{\vk,-}\rangle,
\end{equation}
with the only difference that the occupations $c_{\vk,\pm}(t)$ change over time.
The off-diagonal part of the velocity operator is then
\begin{equation}
  \langle \vop^\ZB \rangle_\vk (t) = 2\operatorname{Re}\lbrace c_{+,\vk}(t) c_{-,\vk}^\ast(t) \, \me^{-\iu \Omega^\st_\vk t } \langle \varphi_{\vk,-}|\vop|\varphi_{\vk,+}\rangle \rbrace \label{eq:generalZB(t)}.
\end{equation}
By definition, the time-dependent occupations can be written as
\begin{equation}
  c_{\vk,\pm}(t) = \langle \varphi_{\vk,\pm} \mid U_I(t,0) \mid \psi_\vk(t=0)\rangle
\end{equation}
with the help of the time-evolution operator in the interaction picture $U_I$, which can be expressed in terms of  a Dyson series \cite{JJSakuraibook},
\begin{align}
 U_I(t) =& 1-\frac{\iu}{\hbar} \int\limits_0^t \; V_I(t^\prime) \dint t^\prime  \nonumber \\
     &+ \left( -\frac{\iu}{\hbar}\right)^2 \int\limits_0^t\dint t^\prime \int\limits_0^{t^\prime} \dint t^{\prime\prime}\, V_I(t^\prime)V_I(t^{\prime\prime}) + \dots{} \, .
\label{eq:DysonSeries-U_I}
\end{align}
With a harmonic driving $V_I\propto \me^{\iu \omega_D t}$, as used in Sec.~\ref{sec:timedep-ZB} the first order of the occupations $c_{\vk,\pm}$ contains terms with the same frequency $\omega_D$.
In order $n$ however, we will get $\me^{\iu n\omega_D t}$ terms of the occupations, which will be reflected in the velocity, see Eq.~\eqref{eq:generalZB(t)}.
Thus, for larger driving amplitudes, higher order oscillations in $\omega_D$ are expected.

\section{Additional long-time ZB modes}
\label{app:crossing_survive}

\begin{figure}
 \centering 
    \def\svgwidth{0.98\columnwidth}
    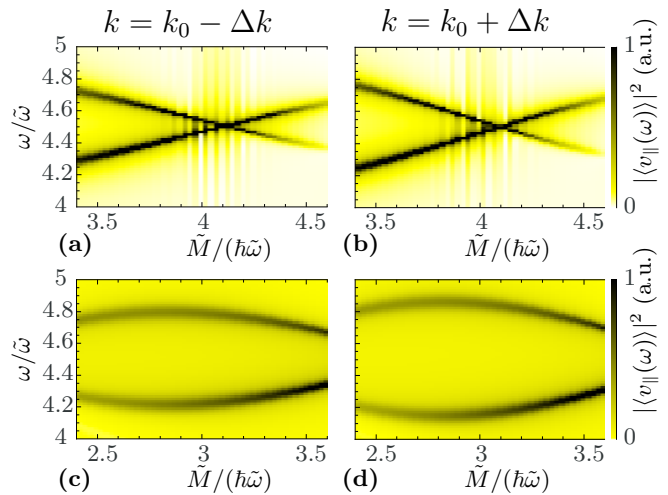
\caption{
Analysis explaining why the modes at the crossing ($\omega\approx n\omega_D$) survive in Fig.~\ref{fig:ZB-FT-longt}(d). 
Here, the parallel velocity $|\langle v_\parallel(\omega)\rangle|^2$ is shown for $M_0=0$ as a function of $\tilde M$ for $\omega_D=1.5\tilde\omega$ with the normalization frequency $\tilde\omega = 2.5 v_F k_0$, for two different energies with $k=0.36\tilde\omega/v_F$ [in (a) and (c)] and $k=0.44\tilde\omega/v_F$ [in (b) and (d)]. The frequency of the ZB is quite independent at the crossings, \ie (a) and (b) are very similar, whereas away from the crossing, it is energy dependent, \ie (c) and (d) differ. 
Since only energy-independent modes survive as discussed in the main text, only modes at the crossings can be seen after at late times [Fig.~\ref{fig:ZB-FT-longt}(d)].
 \label{fig:ZB-FT-Ek-dep}  }
\end{figure}

In this appendix, we discuss numerically the additional (partly) surviving modes shown in Fig.~\ref{fig:ZB-FT-longt}(c) and (d).
There, the modes with $\omega = n \omega_D$ survive, as discussed in the main text, but also the other modes close to the crossings with the horizontal lines $\omega = n \omega_D$. 
Since these are located in regions not explained by our analytical approximations ($\tilde M$ is too large), we can investigate their $\vk$-dependence -- and thus whether they dephase or not over the wave packet's width -- only numerically.
To this end, we simulate the propagation of wave packets with different initial energies for a gapless Dirac cone ($M_0=0$) because the effect can be better seen there.
So far, the propagated Gaussian wave packet is centered around $\vk_0 =  (0.4 M_0/\hbar v_F,0)^T $ with a $\vk$-space width of $\Delta k = k_0/10 $. 
To analyze the $\vk$-dependence of the ZB modes, now we simulate two wave packets centered around $\langle \vk\rangle  = \vk_0 \pm  (\Delta k,0)^T$ instead. 
If the positions of the modes do not change in the diagrams, they are {\it not} $\vk$-dependent (at least for the relevant modes in the wave packet) and are supposed to survive.
On the other hand, if the frequencies of the modes do change, they are $\vk$-dependent and should therefore dephase.
In Fig.~\ref{fig:ZB-FT-Ek-dep}, the data of these additional simulations are shown, \ie wave packets of different initial energies are propagated, in a much smaller parameter regime than before. 
Here, the fixed parameters are $M_0=0$ and $\omega_D =1.5\tilde\omega$, with the unit frequency $\tilde\omega = 2.5 v_F k_0$. 
Panels (a) and (b) show the ZB modes at the crossing of the horizontal line, $\omega= 4.5\tilde \omega = 3 \omega_D$, for different initial energies, whereas panels (c) and (d) depict the region away from those crossings.
Since the position of the crossing at $\tilde M\approx 4.1 \hbar \tilde\omega$ does not considerably change, the crossings are rather $\vk$-\emph{in}dependent over the width of the wave packet.
On the other hand, the frequency change of the modes away from the crossing is visible with the bare eye, since there the $\vk$-dependence is much stronger.
Consequently, only the modes around the crossings are supposed to persist as it is shown in the long-time behavior, where modes in the other regions decay over time.
 
\section{Phase influence of the ZB echo}
\label{app:ZBecho-phases}

This appendix deals with the rather technical point concerning the phase change of the velocity matrix element $\nu_\vk = \langle \varphi_{\vk,+} | v_i | \varphi_{\vk,-} \rangle$ due to the pulse in Eq.~\eqref{eq:v_k^ZBt>t0}.
For gapless Dirac systems, the complex number $\langle \varphi_{\vk,-} | v_i  | \varphi_{\vk,+} \rangle$ is purely imaginary leading to a constant phase jump of $\pi$ due to $\langle \varphi_{\vk,+} | v_i | \varphi_{\vk,-} \rangle = \langle \varphi_{\vk,-} | v_i | \varphi_{\vk,+} \rangle^\ast$ which does not effect the amplitude of the echo. 
For a general system on the other hand, this phase change might be $\vk$-dependent, implying possibly an altered interference of the modes in a wave packet [compare Eq.~\eqref{eq:v^ZBt>t0}].
However, this should in general not lead to a further reduction of the ZB compared to the initial ZB---as long as this phase $\nu_\vk$ is unrelated to the phase of $\alpha_\vk = c_\vk^+(c_\vk^-)^\ast$.
In this case, the interference of different $\vk$-modes at the start and at the echo time is expected to be qualitatively the same.

An exception would be the case $\nu_\vk = \alpha_\vk$ for all $\vk$. 
Note that this is highly unlikely because these two quantities are independent---$\alpha_\vk$ depends only on the initial wave packet, whereas $\nu_\vk$ only on unperturbed Hamiltonian and the velocity operator.
In this unrealistic correlated case, the ZB at $t=0$ would be diminished because of nonperfect interference of different $\vk$-modes. 
On the other hand, the ZB at $t_1 = t_0$ would be enhanced since $\alpha_\vk$ and $-\nu_\vk$ would exactly cancel for all $\vk$ leading to perfect interference of all modes. 
Thus for a transition amplitude close to one ($|A(\vk,\Delta t)|\lesssim 1$), the amplitude of the echo ZB could be even higher than of the initial ZB.
However, we do not consider this highly unlikely case further in this paper.
Instead, we neglect the effect of $\nu_\vk$, which is justified for narrow wave packets in $\vk$-space such that $\nu_\vk\simeq\nu_{\vk_0}$.

\section{Effect of disorder on the ZB echo}
\label{app:disorderZB}

\begin{figure}
 \centering 
\includegraphics[width=\columnwidth]{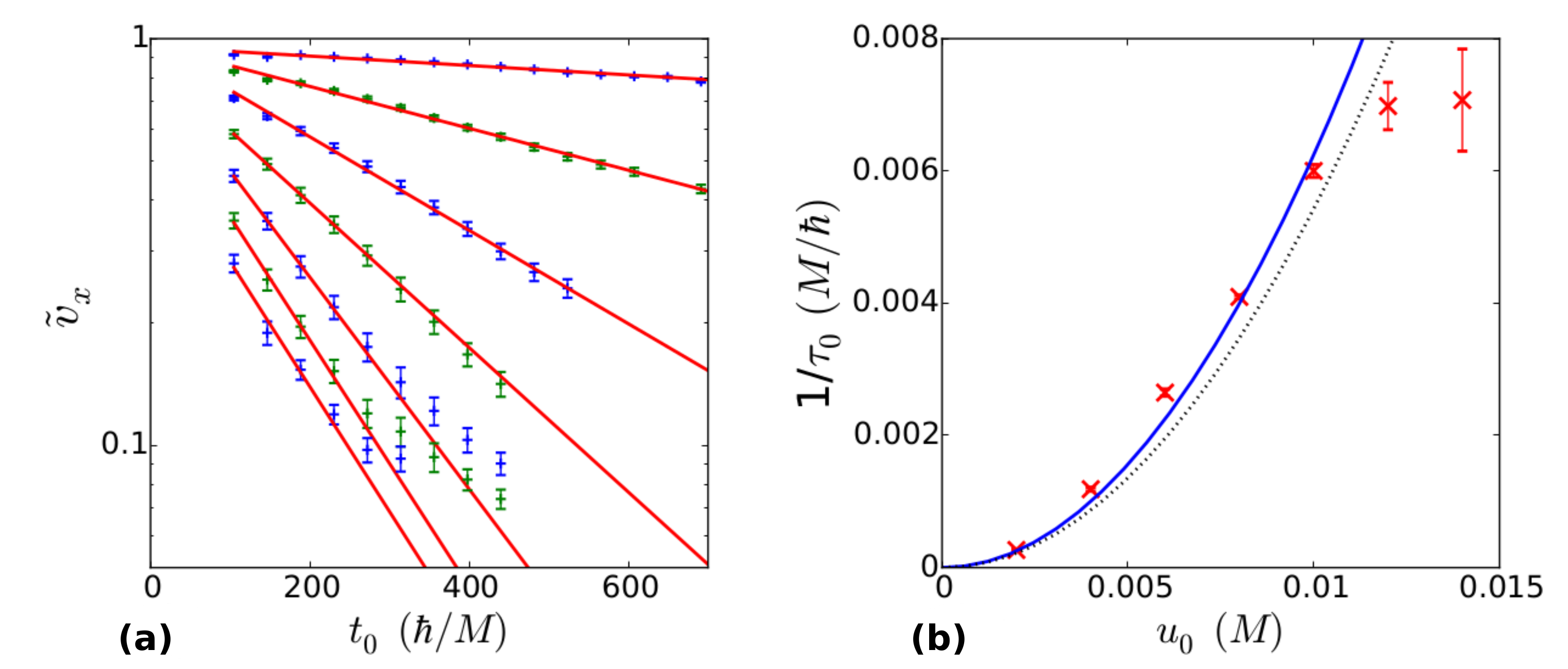}
\caption{ 
ZB echoes in presence of disorder. 
(a) The time-dependent revival strength $\tilde v_x=B_x^\revival/B_x^\initial$ is shown as a function of the pulse time $t_0$ for different disorder strengths ranging from $u_0 = 0.002 M$ to $0.014M$. 
An exponential decay can be seen in the weak disorder regime. 
(b) The decay rate $1/\tau_0$ is extracted by a fit [red lines in (a)] and plotted as a function of $u_0$, and compared to the analytically expected scattering time (black dotted) from Eq.~\eqref{eq:elscattime_result}. 
The quadratic fit in $u_0$ (blue line) to the data points is close to the expected scattering time. 
Saturation is obtained for larger $u_0$. For more details, see the discussion of disorder in Ref.~\cite{reck2017}.
} \label{fig:ZB-revival-disorder}
\end{figure}

In the main text, we showed that the echo of ZB induced by our QTM behaves as expected in pristine and gapped Dirac systems. 
Now, we could continue and verify all other results obtained in the general paper about QTMs \cite{reck2018b}, such as asymmetric band structures and position-dependent potentials in the Hamiltonian.
There is no reason for the results to differ from the QTM case, and hence why we only show exemplarily the effect of disorder discussed in Refs.~\cite{reck2017, reck2018b}.
The previous results show that disorder cannot be effectively time-inverted by the QTM and leads to an exponential decay of the echo strength (measured by the echo fidelity) as function of propagation time, which is in the present case the echo time $t_\echo \simeq 2t_0$.
Due to the overlap of states with positive and negative energy in the ZB, the relative amplitude of the velocity is closely related to the echo fidelity and we expect a similar behavior, \ie an exponential decay.
	
To compare with the previous results, we use the same setup as in Ref.~\cite{reck2017}, \ie pristine graphene with a pulse that opens a mass gap of strength $\tilde M = M$ and an pseudospin-independent impurity potential $V_\imp(\vq)$.
The only difference compared to Ref.~\cite{reck2017} is that, in order to generate ZB, the initial wave packet lives in both bands. 

In order to generate the random, inhomogeneous disorder potential $V_\text{imp}$, every grid point is assigned a normal distributed random number $\beta_i$. 
To avoid a highly fluctuating potential, an average over neighboring points weighted by a Gaussian profile with a range $l_0$, is taken at each site:
\begin{equation}
 V_{\text{imp}}(\vq) = \frac{u_{0}}{\mathcal{N}} \sum\limits_{i=1}^{N} \beta_i \me^{-\frac{(\vq-\vq_i)^2}{l_0^2}} \label{eq:disorderpot}.
\end{equation}
Here, the sum runs over grid points, $u_0$ is related to the mean impurity strength and $\mathcal{N}$ is a normalization factor to account for different realizations with the same parameters $u_0$ and $l_0$,
\begin{equation}
 \mathcal{N} = \left[\frac{1}{A}  \int_A \,{\rm d}^2r \left( \sum\limits_{i=1}^{N} \beta_i \me^{-\frac{(\vq-\vq_i)^2}{l_0^2}} \right)^2 \right]^\frac{1}{2}, 
\label{eq:disorderNormalize}
\end{equation}
where $A$ is the finite area of the grid. 
$\mathcal{N}$ can be thought of the mean deviation of the potential strength over the whole area.

The elastic scattering time $\tau_0$ linked to $ V_{\text{imp}}(\vq)$ is given by (see \eg Ref.~\cite{akkermansbook}):
\begin{equation}
 \frac{\hbar}{\tau_0} = \frac{2\pi}{v_F\hbar} k \,u_0^2 l_0^2 \exp\left(-k^2 l_0^2\right)  I_0\left(k^2 l_0^2\right), 
 \label{eq:elscattime_result}
\end{equation}
where $I_0$ is the modified Bessel function of $0$-th kind.
%
%

As before, we measure the revival of the ZB through the relative echo amplitude, 
$B_i^\revival/B_i^\initial$,
which is plotted as a function of the pulse time $t_0$ for different disorder strengths $u_0$ in Fig.~\ref{fig:ZB-revival-disorder}(a).
Indeed, we see exponential decays of the echo as function of the pulse time $t_0$, where again a saturation is achieved for high $u_0$ and $t_0$, in analogy to  Ref.~\cite{reck2017}.	
The fitted decay rates $1/\tau_0$ of panel (a) (red lines) are plotted in panel (b) as function of the disorder strength $u_0$, which are expected to increase quadratically (compare Eq.~\eqref{eq:elscattime_result}).
Up to some value of $u_0\simeq0.012 M$, the decay rate indeed increases quadratically, as can be seen by the quadratic fit (blue) and is slightly larger but close enough to the purely analytically expected decay (black dotted line).
Above $u_0\simeq0.012 M$, deviations are expected \cite{reck2017} since the golden-rule-decay regime is no longer valid, which was used to derive $\tau_0$ in Eq.~\eqref{eq:elscattime_result}. 

\bibliographystyle{apsrev4-1}
\bibliography{ZB_biblio}


\end{document}